\documentclass[aps,english,notitlepage,nofootinbib]{revtex4-1}

\usepackage{amstext,amsmath,amssymb,amsfonts,bbm}
\usepackage{epsfig}
\usepackage{hyperref}
\usepackage{amsthm}
\usepackage{subfig}
\usepackage{wrapfig}
\usepackage{color}
\usepackage{multirow}
\usepackage{soul}
\usepackage{epigraph}
\usepackage{csquotes}

\usepackage[latin9]{inputenc}
\usepackage{babel}
\usepackage{enumitem}

\usepackage[font=small,labelfont=bf,
   justification=raggedright,
   format=plain]{caption}

\usepackage{titlesec}
\titleformat*{\paragraph}{\bf\small}

\usepackage{array}
\usepackage{mathtools}
\usepackage{amsmath}
\usepackage{amsthm}
\usepackage{amssymb}
\usepackage{cancel}

\theoremstyle{plain}

\theoremstyle{plain}

\theoremstyle{definition}

\theoremstyle{plain}

\theoremstyle{plain}
\newtheorem*{lem*}{\protect\lemmaname}
\theoremstyle{plain}
\newtheorem*{cor*}{\protect\corollaryname}
\theoremstyle{plain}
\newtheorem*{thm*}{\protect\theoremname}
\theoremstyle{plain}

\providecommand{\corollaryname}{Corollary}
\providecommand{\definitionname}{Definition}
\providecommand{\lemmaname}{Lemma}

\providecommand{\theoremname}{Theorem}

\usepackage{import}
\usepackage{tikz}
\usetikzlibrary{tikzmark}

\newcommand{\bea}{\begin{eqnarray}}
\newcommand{\eea}{\end{eqnarray}}

\newcommand{\be}{\begin{equation}}
\newcommand{\ee}{\end{equation}}

\begin{document}
\begin{flushleft}
KCL-PH-TH/2019-41
\par\end{flushleft}

\date{\today}

\title{Group field theory condensate cosmology: An appetizer}

\author{Andreas G. A. Pithis}
\email{andreas.pithis@kcl.ac.uk}


\author{Mairi Sakellariadou}
\email{mairi.sakellariadou@kcl.ac.uk}

\affiliation{Department of Physics, King's College London, University of London, \\
Strand WC2R 2LS, London, United Kingdom (still E.U.)}

\begin{abstract}

This contribution is an appetizer to the relatively young and fast evolving approach to quantum cosmology based on group field theory condensate states. We summarize the main assumptions and pillars of this approach which has revealed new perspectives on the long-standing question of how to recover the continuum from discrete geometric building blocks. Among others, we give a snapshot of recent work on isotropic cosmological solutions exhibiting an accelerated expansion, a bounce where anisotropies are shown to be under control and inhomogeneities with an approximately scale-invariant power spectrum. Finally, we point to open issues in the condensate cosmology approach.

\end{abstract}

\keywords{}

\pacs{}
\maketitle

\epigraph{Most important part of doing physics is the knowledge of \textit{approximation}.}{Lev Davidovich Landau}

\section{Introduction}\label{sec:intro}

The current observational evidence strongly suggests that our Universe is accurately described by the standard model of cosmology~\cite{Planck2018}. This model relies on Einstein's theory of general relativity (GR) and assumes its validity on all scales. However, this picture proves fully inadequate to describe the earliest stages of our Universe, as our concepts of spacetime and its geometry as given by GR are then expected to break down due to the extreme physical conditions encountered in the vicinity of and shortly after the big bang. More explicitly, it appears that our Universe emerged from a singularity, as implied by the famous theorems of Penrose and Hawking~\cite{Singularitytheorems}. From a fundamental point of view, such a singularity is unphysical and it is expected that quantum effects lead to its resolution~\cite{deWitt}. This motivates the development of a quantum theory of gravity in which the quintessential features of GR and quantum field theory (QFT) are consistently unified. Such a theory will revolutionize our understanding of spacetime and gravity at a microscopic level and should be able to give a complete and consistent picture of cosmic evolution. 

The difficulty in making progress in this field is ultimately rooted in the lack of experiments which have access to the physics at the smallest length scales and highest energies and so would provide a clear empirical guideline for the construction of such a theory. In turn, this severe underdetermination of theory by experiment is a reason for the current presence of a plethora of contesting approaches to quantum gravity~\cite{approaches}. Against this backdrop, the cosmology of the very early Universe represents a unique window of opportunity out of this impasse~\cite{qgcintimate}. For instance, it is naturally expected that traces of quantum gravity have left a fingerprint on the spectrum of the cosmic microwave background radiation, see e.g. Refs.~\cite{qgimprint}. Hence, cosmology provides an ideal testbed where the predictions of such competing theories can be compared and tested against forthcoming cosmological data.

A central conviction of \textit{some} approaches to quantum gravity is that it should be a non-perturbative, background-independent and diffeomorphism invariant theory of quantum geometry. In this sense, the spacetime continuum is renunciated and is instead replaced by degrees of freedom of discrete and combinatorial nature.\footnote{The introduction of discrete structures can be motivated to bypass the issue of perturbative non-renormalizability of GR within the continuum path integral formulation. Alternative points of view of dealing with this issue would be to assume the existence of a non-perturbative (i.e. interacting) fixed point for gravity in the UV as done by the asymptotic safety program~\cite{ASG} or to increase the amount of symmetries as compared to GR and QFT with the aim to regain perturbative renormalizability as proposed by string theory~\cite{ST}. Yet another view, as presented by non-commutative geometry, is that above the Planck scale the concept of geometry collapses and spacetime is replaced by a non-commutative manifold~\cite{NCG}.} Particular representatives of this class of theories are the closely related canonical and covariant loop quantum gravity (LQG)~\cite{LQG,covlqg}, group field theory (GFT)~\cite{GFT}, tensor models (TM)~\cite{TM} and simplicial quantum gravity approaches like quantum Regge calculus (QRC)~\cite{QRC} and Euclidean and causal dynamical triangulations (EDT, CDT)~\cite{EDTCDT}. The perturbative expansion of their path integrals each yields a sum over discrete geometries and the most difficult problem for all of them then lies in the recovery of continuous spacetime geometry and GR describing its dynamics in an appropriate limit. This is challenging because it ideally requires to formulate statements about the continuum by only calling upon notions rooted in the discontinuum. Taking the continuum limit in these approaches, crucially depends on whether the discreteness of geometry is considered physical or unphysical therein, the proper weighting of configurations in the partition function and the precise specification of the continuum limit itself. Consequently, strategies to reach this goal differ among them strongly, see e.g. Ref.~\cite{thesis} for an overview.

In the light of the above, it is vitally important to consider these approaches in a cosmological context which has been accomplished to a varying degree of success by them. In this contribution we give a brief and rather non-technical panorama of the GFT condensate cosmology program~\cite{GFC} which has been developed over the last few years and has so far borne promising fruits.\footnote{For previous articles giving a review account of the program, we refer to Refs.~\cite{GFCreview,Disappearance2}.} This program is motivated by the idea that the mechanism for regaining a continuum geometry from a physically discrete quantum gravity substratum in GFT is provided by a phase transition to a condensate phase~\cite{Disappearance,Disappearance2}. Research on the phase structure of different GFT models in terms of functional renormalization group analyses finds support for such a conjecture in terms of IR fixed points~\cite{TGFTFRG1,TGFTFRG2}. Further backing is provided by saddle point studies~\cite{GFTmin} and Landau-Ginzburg mean field analyses~\cite{GFTLandau} which probe non-perturbative aspects of GFT models, see also Ref.~\cite{thesis}. In this picture, a condensate would correspond to a non-perturbative vacuum which comprises of a large number of bosonic GFT quanta and in the context of GFT models of four-dimensional quantum gravity is tentatively interpreted as a continuum geometry. Given this basic premise, the most striking successes and milestone results of the condensate cosmology approach are the recovery of Friedmann-like dynamics of an emergent homogeneous and isotropic geometry~\cite{GFCFriedmann}, an extended accelerated phase of expansion~\cite{GFCaccel1,GFCaccel2} right after a cosmological bounce~\cite{GFCbounce}, a simple yet effective mechanism for dynamical isotropization of microscopic anisotropies~\cite{aniso1,aniso2} and the finding of an approximately scale-invariant and small-amplitude power spectrum of quantum fluctuations of the local volume over a homogeneous background geometry perturbed by small inhomogeneities~\cite{GFCinhom}. In the following, we will quickly review the GFT formalism, give the basic structures behind its condensate cosmology spin-off, highlight the main results and give an outlook for future challenges of this program.\footnote{Another way to relate GFT to cosmology was brought forward in Ref.~\cite{CalcagniGielenOriti}. This work is closer to canonical quantum cosmology (either Wheeler-DeWitt or loop quantum cosmology) in the sense that it is built on a minisuperspace model, i.e., symmetry reduction is applied before quantization and not afterwards as in the condensate program.}

\section{Group field theory}\label{sec:GFT}

GFTs are quantum field theories which live on group configuration spaces, possess a gauge symmetry and are in particular characterized by combinatorially non-local interactions~\cite{GFT}. More precisely, the real- or complex-valued scalar field $\varphi$ lives on $d$ copies of a Lie group $G$. In models for quantum gravity, $G$ corresponds to the local gauge group of GR and the gauge symmetry leads to an invariance of the GFT action under the (right) diagonal action of $G$ which acts on the fields as
\be\label{invariance}
\varphi(g_1,...,g_d)=\varphi(g_1 h,...,g_d h),~\forall g_i,h\in G.
\ee
For $4d$ quantum gravity models $G$ is typically $\textrm{SO}(3,1)$ (or $\textrm{SL}(2,\mathbb{C})$) in the Lorentzian case, $\text{SO}(4)$ (or $\textrm{Spin}(4)$) in the Riemannian case or their rotation subgroup $\textrm{SU}(2)$ which is the gauge group of Ashtekar-Barbero gravity. The group  elements $g_I$ with $I = 1,...,d$ are parallel transports $\mathcal{P} \text{e}^{i\int_{e_{I}}A}$ which are associated to $d$ links $e_I$ and $A$ denotes a gravitational connection $1$-form. The gauge symmetry guarantees the closure of the faces dual to the links $e_I$ to form a $d-1$-simplex. For the most discussed case where $d=4$, one obtains tetrahedra in this way. The metric information encoded in the fields can be retrieved via a non-commutative Fourier transform~\cite{NCFT,GFC,GFCreview,GFCExample,aniso1}.

For a complex-valued field the action has the structure
\be
S[\varphi,\bar{\varphi}]=\int (\text{d}g)^d\bar{\varphi}(g_I)\mathcal{K}(g_I)\varphi(g_I)+\mathcal{V}[\varphi,\bar{\varphi}],
\ee
where we used the shorthand notation $\varphi(g_I)\equiv \varphi(g_1,...,g_d)$. Since further details about the action are specified below, here the following suffices say. The local kinetic term typically incorporates a Laplacian and a \enquote{mass term} contribution. The former is  motivated by renormalization studies on GFT~\cite{GFTren} while the latter can be related to spin foam edge weights via the GFT/spin foam correspondence~\cite{GFTSF} (see below) and hence should not be confused with a physical mass. The so-called simplicial interaction term consist of products of fields paired via convolution according to a combinatorial non-local pattern which for $d=4$ encodes the combinatorics of a $4$-simplex. The precise details of the kinetic and interaction term are supposed to encode the Euclidean or Lorentzian embeddings of the theory~\cite{GFCFriedmann,aniso2,GFTLorentzian}. 

With this, the perturbative expansion of the parition function 
\be
Z_{\text{GFT}}=\int[\mathcal{D}\varphi][\mathcal{D}\bar{\varphi}]
\text{e}^{-S[\varphi,\bar{\varphi}]}
\ee
is indexed by Feynman diagrams which are dual to gluings of $d$-simplices.\footnote{Notice that by attributing an additional combinatorial degree of freedom named color to the fields, one can guarantee that the terms of the perturbative expansion are free of topological pathologies~\cite{coloring}.} In this way, it provides a generating function for the covariant quantization of LQG in terms of spin foam models~\cite{covlqg}. In LQG, boundary spin network states of a spin foam correspond to $3$-dimensional discrete quantum geometries while the spin foam transition amplitudes interpolate in between two such boundary configurations. There, a proper imposition of the so-called simplicity constraints guarantees that $\text{SL}(2,\mathbb{C})$- or $\text{Spin}(4)$-data in the bulk is reduced to $\text{SU}(2)$-valued data on the boundary~\cite{SFSC,GFTEPRLEuclidean,GFTSG}. The main aspects of the GFT formulation of the currently most studied spin foam model for Lorentzian 4d quantum gravity, the so-called EPRL model~\cite{covlqg} are specified by the aforementioned simplicial interaction term, that the GFT fields are defined over $\text{SU}(2)^4$ (thus encoding the boundary geometry) and finally the proper embedding of these data into $\text{SL}(2,\mathbb{C})$ which is realized by the dynamics and thus is encoded by the details of the kinetic and interaction term in the action. We further specify this action in Sections~\ref{condensatedynamics} and~\ref{subsection:resultsA} but note here that all the details of the interaction term have so far not been put down in terms of its boundary data~\cite{aniso2} which, for what this review is concerned, does not pose a limitation.

In the second quantized formulation of GFT, introduced in Ref.~\cite{GFTLQG}, motivated by the origins of GFT in LQG, spin network boundary states are viewed as elements of the GFT Fock space wherein spin network vertices, i.e. atoms of space, correspond to fundamental quanta which are created or annihilated by the field operators of GFT.\footnote{A detailed discussion on the subtle differences in between the Fock space of GFT and the kinematical Hilbert space of LQG, which are mostly related to the absence of the so-called cylindrical consistency and equivalence in the former, is found in Ref.~\cite{GFTLQG}.} The GFT Fock space
\be
\mathcal{F}(\mathcal{H}_v)=\bigoplus_{N=0}^{\infty}\textrm{sym}\bigl(\otimes_{i=1}^N\mathcal{H}_v^{(i)}\bigr),
\ee
is built by means of the fundamental Hilbert space $\mathcal{H}_v=L^2(G^d)$ of a GFT quantum which is assumed to obey bosonic statistics.\footnote{The assumption of bosonic statistics is crucial for the condensate cosmology program where spacetime is thought to arise from a GFT condensate. To justify this choice of statistics from a fundamental point of view is an open problem, see Ref.~\cite{GFTLQG} for a discussion and Refs.~\cite{coloring,GFTotherstat} for explorations into other statistics.} Clearly, for $G=\textrm{SU}(2)$ and imposing gauge invariance as in Eq. (\ref{invariance}), a state in $\mathcal{H}_v$ represents an open LQG spin network vertex or its dual quantum polyhedron. In particular, for $d=4$ a GFT quantum corresponds to a quantum tetrahedron which also is the most studied case within the condensate cosmology program~\cite{GFC,GFCreview}. In the remainder, we stick to this choice for $G$ and $d$.

In this picture, many particle GFT states can be excited over the Fock vacuum $|\emptyset\rangle$ which is the state devoid of any topological and quantum geometric information. Standardly, it is defined via the action of an annihilation field operator, namely
\be
\hat{\varphi}(g_I)|\emptyset\rangle=0,
\ee
where the vacuum is normalized to $1$. Given their bosonic statistics, the GFT field operators obey the canonical commutation relations
\be
\bigl[\hat{\varphi}(g_I),\hat{\varphi}^{\dagger}(g_I')\bigr]=\int \text{d}h\prod_{I}\delta(g_I h g'^{-1}_{I})~\textrm{and}~\bigl[\hat{\varphi}^{(\dagger)}(g_I),\hat{\varphi}^{(\dagger)}(g_I')\bigr]=0,
\ee
where the form of the delta distribution accounts for the imposition of gauge invariance, Eq. (\ref{invariance}).

In this framework, quantum geometric observable data can be retrieved from such states via second-quantized Hermitian operators~\cite{GFTOperators}, e.g. the number operator is given by
\be
\hat{N}=\int (\text{d}g)^d \hat{\varphi}^{\dagger}(g_I)\hat{\varphi}(g_{I})
\ee
while more general one-body operators read as
\be
\hat{O}=\int (\text{d}g)^d \int (\text{d}g')^d~\hat{\varphi}^{\dagger}(g_I)O(g_I,g_I')\hat{\varphi}(g_{I}'),
\ee
wherein $O(g_I,g_I')$ denote the matrix elements of a corresponding first-quantized operator. In this way, the area and volume operator of LQG can be imported into the GFT context, which is typically done by working in the spin representation introduced below.\footnote{We refer e.g. to Appendix C of Ref.~\cite{aniso2} for an extensive discussion of this matter for the case of the volume operator.} Hence, in GFT the discreteness of geometry is considered as being real, rooted in its strong connections to LQG where the spectra of geometric operators are discrete (as shown to hold at the kinematical level)~\cite{LQGdiscreteness}.

\section{Group field theory condensate cosmology}
\label{sec:GFTCC}

The general aim of the GFT condensate cosmology program is to describe cosmologically relevant geometries by means of the the formalism given above. Concretely, the goal is to \textit{approximate} $3$-dimensional homogeneous and extended geometries as well as their cosmological evolution in terms of GFT condensate states and their effective dynamics.

\subsection{Motivation for condensate states}\label{condensatestates}

\begin{figure}[!h]
\centering%
\begin{minipage}{0.5\textwidth}
  \includegraphics[width=0.85\linewidth]{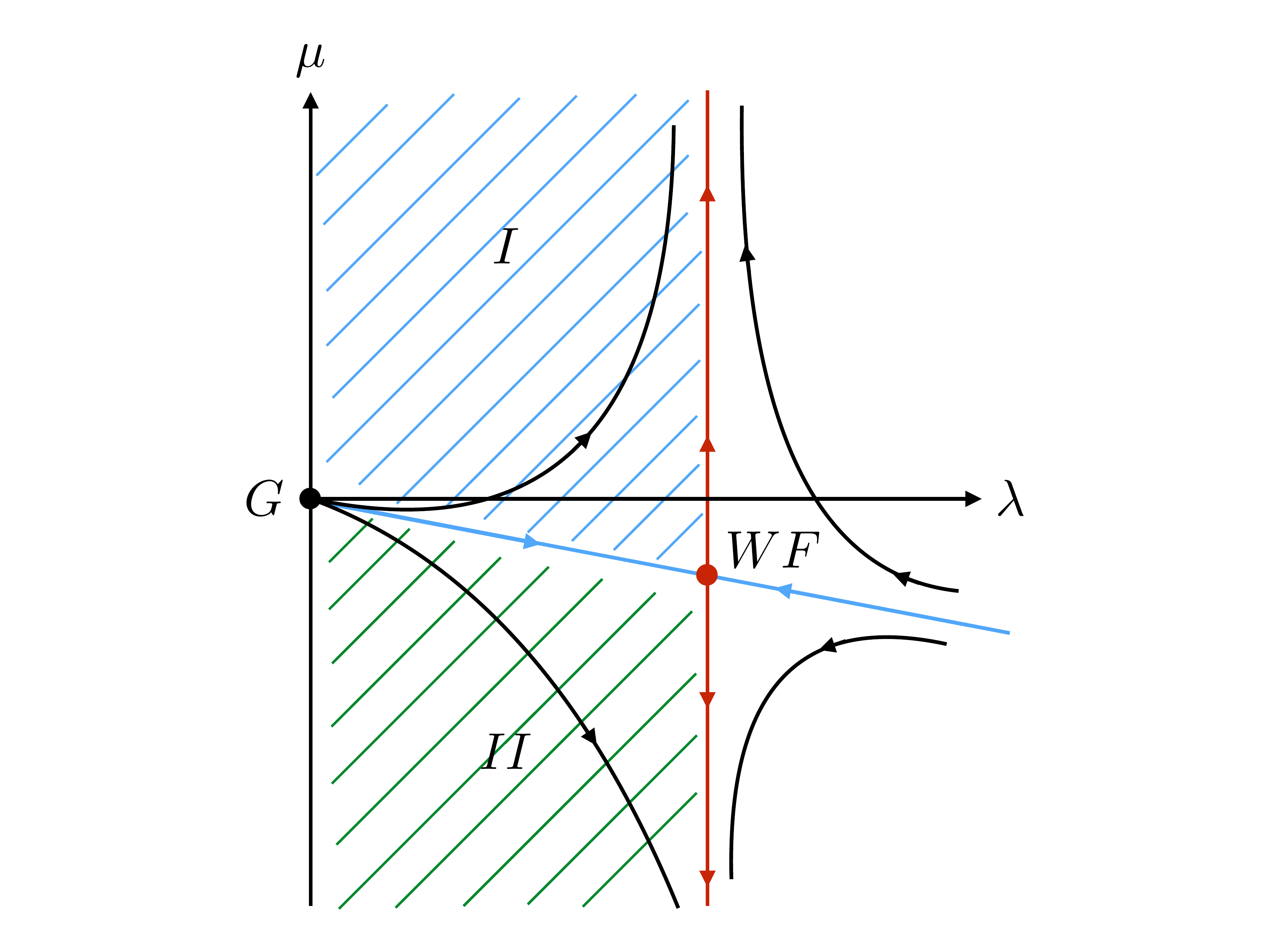}
\end{minipage}%
\begin{minipage}{0.5\textwidth}
  \includegraphics[width=0.85\linewidth]{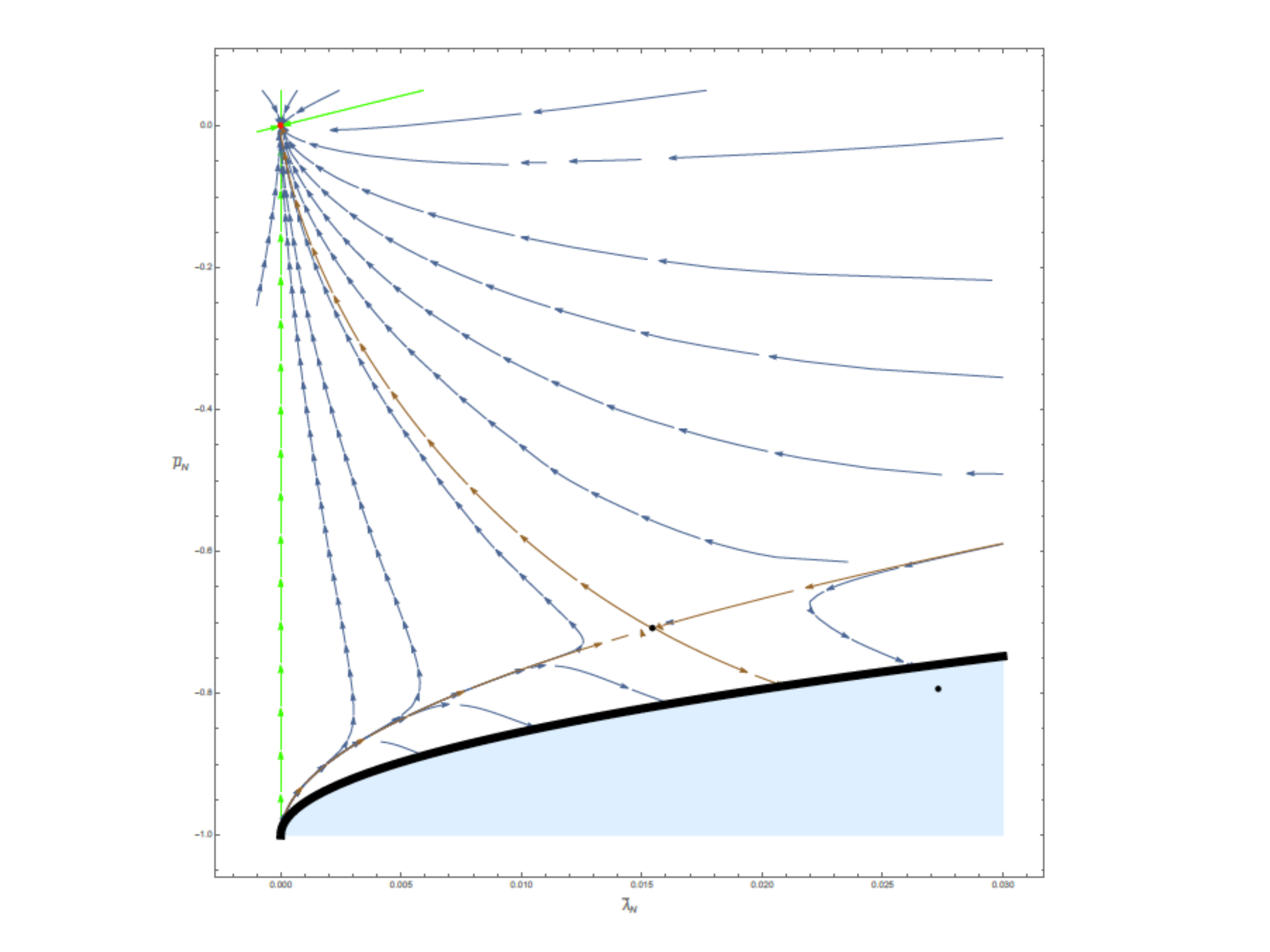}
\end{minipage}\caption[A phase diagram with Wilson-Fisher fixed point and a phase diagram for a tensorial GFT.]{Left: Phase diagram of a local scalar field theory with quartic interaction on $\mathbb{R}^3$. The \enquote{mass parameter} is denoted by $\mu$ while the interaction couples with $\lambda$. $G$ denotes the Gaussian fixed point and $WF$ the Wilson-Fisher fixed point. In the region hatched in green $\langle\hat{\varphi}\rangle\neq 0$ holds. Right: Exemplary phase diagram of a quartic tensorial GFT on $\mathbb{R}^d$ (cf. Refs.~\cite{TGFTFRG1b}). In the analogue of region $II$ on the left hand side, a non-vanishing expectation value of the field operator is expected to be found.}\label{figfixedpoint}
\end{figure}

As initially stated in the introduction, indications for the formation of a condensate phase have been found through the analyses of non-perturbative aspects of GFT models. In particular, functional renormalization group analyses of so-called tensorial GFTs~\cite{TGFTFRG1,TGFTFRG1b} indicate a phase transition separating a symmetric from a broken/condensate phase as the \enquote{mass parameter} tends to negative values in the IR limit which is analogous to a Wilson-Fisher fixed point in the corresponding local QFT. This is illustrated in terms of the phase diagrams in Fig.~\ref{figfixedpoint}. Building on these, more work has to be devoted to studying the phase structure of a (potentially coloured) GFT model enriched with additional geometric data and an available simplicial quantum gravity interpretation. The hope would be that in the phase diagram of such a theory at least one phase can be found which can be interpreted as a physical continuum geometry of relevance to cosmology.\footnote{Complementarily to the application of functional methods to study the notion of phases in this context, research on the algebraic foundations of GFT has shown the existence of representations which are unitarily inequivalent to the one of the GFT Fock space and that are potentially related to different phases of GFT models, in particular to condensate phases~\cite{GFCstatic,GFTinequiv,GFTequilib}.}

Given this central hypothesis of the GFT condensate cosmology program, the goal is to directly derive the effective dynamics for GFT condensate states from the microscopic quantum dynamics using mean field techniques inspired by the theory on Bose-Einstein condensates~\cite{BECs} and to extract a cosmological interpretation thereafter. Generally, a condensate phase corresponds to a non-perturbative vacuum of a theory where the expectation value of the field operator is non-vanishing, i.e. $\langle \varphi(g_I)\rangle\neq 0$. Since such a vacuum is described by a large number $N$ of quanta, in the GFT context this would make it suitable to model extended geometries. In addition, these quanta are occupying the same quantum geometric configuration which is desirable if a homogeneous background geometry is to emerge from the condensate. Simple trial states which capture these features are field coherent states of the form
\be
|\sigma\rangle=A~e^{\hat{\sigma}}|\emptyset\rangle,~~~\hat{\sigma}=\int (\text{d}g)^d~\sigma(g_I)\hat{\varphi}^{\dagger}(g_I)~~~\text{and}~~~A=\text{e}^{-\frac{1}{2}\int(\text{d}g)^d~|\sigma(g_I)|^2}
\ee
corresponding to an infinite superposition of states which for $d=4$ describe disconnected quantum tetrahedra labeled by the same discrete geometric data. The latter is encoded by a single collective function, the condensate wave function $\sigma$. These states are field coherent since they are eigenstates of the field operator,
\be
\hat{\varphi}(g_I)|\sigma\rangle=\sigma(g_I)|\sigma\rangle,
\ee
for which $\langle\hat{\varphi}(g_I)\rangle=\sigma(g_I)\neq 0$ holds. Finally, in addition to the right invariance as in Eq. (\ref{invariance}), we require the invariance under the left diagonal action of $G$, i.e., $\sigma(k g_I)=\sigma(g_I)$ for all $k\in G$ to guarantee that the domain of the condensate wave function is isomorphic to the minisuperspace of homogeneous geometries $\text{GL}(3,\mathbb{R})/\text{O}(3)$~\cite{GFCExample}.\footnote{In principle, more complex composite states can be constructed so as to encode connectivity information in between GFT quanta and topological information to model e.g. spherical geometries~\cite{GFTOperators,GFTBH}.}\footnote{One may also take the view that the existence of a condensate phase transition is of less pronounced importance for such condensate states to be suitable non-perturbative states of physical relevance. We refer to Ref.~\cite{GFCuniverse} for a detailed discussion.}

\subsection{Effective condensate dynamics}\label{condensatedynamics}

An effective dynamics of such states can be obtained by taking alternative but equivalent roads. One can either study the GFT path integral in saddle point approximation or use the lowest-order truncation of the Schwinger-Dyson equations of the GFT model under consideration~\cite{GFC,GFCreview,rest,GFTmin,GFTLandau}. These equations can be derived when using 
\be
0=\delta_{\bar{\varphi}}\langle\mathcal{O}[\varphi,\bar{\varphi}]\rangle=\int[\mathcal{D}\varphi][\mathcal{D}\bar{\varphi}]\frac{\delta}{\delta\bar{\varphi}(g_I)}\left(\mathcal{O}[\varphi,\bar{\varphi}]\text{e}^{-S[\varphi,\bar{\varphi}]}\right)=\left\langle\frac{\delta\mathcal{O}[\varphi,\bar{\varphi}]}{\delta\bar{\varphi}(g_I)}-\mathcal{O}[\varphi,\bar{\varphi}]\frac{\delta S[\varphi,\bar{\varphi}]}{\delta\bar{\varphi}(g_I)}\right\rangle,
\ee
where $\mathcal{O}$ is a functional of the fields. An expression encoding the effective dynamics is then extracted by setting $\mathcal{O}$ to the identity, giving
\be
\left\langle\frac{\delta S[\varphi,\bar{\varphi}]}{\delta\bar{\varphi}(g_I)}\right\rangle=0.
\ee
If we evaluate the expectation value with respect to the condensate state, one yields
\be
\mathcal{K}(g_I)\sigma(g_I)+\frac{\delta \mathcal{V}}{\delta\bar{\sigma}(g_I)}=0
\ee
which is the classical equation of motion for the condensate wave function. Its solution would amount to solving the theory at tree-level. In general, this is a non-linear and non-local equation for the dynamics of the mean field $\sigma$ and is given the interpretation of a quantum cosmology equation despite the fact that it has no  direct  probabilistic  interpretation as compared to the equations of motion of Wheeler-DeWitt (WdW) quantum cosmology~\cite{WdWQC} and loop quantum cosmology (LQC)~\cite{LQC}. However, this does not pose a problem in order to extract cosmological predictions from the full theory, as we will review below.

In a next step, in order to extract information regarding the dynamics of such condensate systems, we extend the set of degrees of freedom of the formalism and couple a free, massless, minimally coupled real-valued scalar field to the GFT field,
\be 
\sigma : G^d \times \mathbb{R}\to \mathbb{C}~~(\text{or}~\mathbb{R}).
\ee
This scalar field serves as a relational clock, i.e. an internal time variable, with respect to which the latter evolves. Such a procedure is common practice in classical and quantum gravity~\cite{partialobs,rods,LQC,dustlqg}. Notice that the expectation values of the above-introduced observables will then obviously depend on the relational clock $\phi$. The precise introduction of this degree of freedom is based on the expression of the Feynman amplitudes of a given simplicial GFT model which take the form of simplicial gravity path integrals for gravity when coupled to such a scalar field, as explained in detail in Refs.~\cite{GFCFriedmann,GFTclock}. In this discretized setting, the matter field sits on the vertices which are dual to the $4$-simplices of the simplicial complex. 

In this way, the action takes the general form
\be\label{actioneprl}
S[\sigma,\bar{\sigma}]=\int (\text{d}g)^d\text{d}\phi~\bar{\sigma}(g_I,\phi)\mathcal{K}(g_I,\phi)\sigma(g_I,\phi)+\mathcal{V}[\sigma,\bar{\sigma}]
\ee
where $\mathcal{K}$ is local in $g_I$ and $\phi$ and the interaction term is given by 
\be
\mathcal{V}[\sigma,\bar{\sigma}]=\frac{\lambda}{5}\int\left(\prod_{a=1}^5\text{d}g_{I_a}\sigma(g_{I_a},\phi)\right)\mathcal{V}_5~~+~~\text{c.c.},
\ee
where each $g_I$ corresponds to four group elements. The object $\mathcal{V}_5=\mathcal{V}_5(g_{I_1},g_{I_2},g_{I_3},g_{I_4},g_{I_5})$ is a function of all group elements, encoding the combinatorics of a $4$-simplex, which when appropriately specified together with $\mathcal{K}$ are supposed to yield the GFT formulation of the EPRL spin foam model for Lorentzian quantum gravity in $4d$~\cite{GFTEPRLEuclidean,GFCFriedmann,aniso2}.

For the remainder of this review it is important to introduce the spin representation of GFT fields. For left- and right-invariant configurations as considered in the context of the condensate program, we may give the Peter-Weyl decomposition of the condensate field as
\be\label{meanfielddecomposed}
\sigma(g_1,g_2,g_3,g_4,\phi)=\sum\limits_{\substack{j_1,...,j_4 \\ m_1,....,m_4\\ n_1,...,n_4\\ \iota_l,\iota_r}}\sigma^{j_1 j_2 j_3 j_4, \iota_l\iota_r}(\phi)\bar{\mathcal{I}}^{j_1 j_2 j_3 j_4,\iota_l}_{m_1 m_2 m_3 m_4}\mathcal{I}^{j_1 j_2 j_3 j_4,\iota_r}_{n_1 n_2 n_3 n_4}\prod_{i=1}^4 d_{j_i} D^{j_{i}}_{m_i n_i}(g_i),
\ee
where $D^{j}_{m n}(g)$ are the Wigner matrices and $d_j=2j+1$ is the  dimension of the corresponding irreducible representation. The  representation  label $j$ is an element of the set $\{0,1/2,1,3/2,...\}$ while the indices $m,n$ assume the values $-j\leq m,n \leq j$. The objects $\mathcal{I}^{j_1 j_2 j_3 j_4,\iota}$ are called intertwiners and are elements of the Hilbert space of states of a single tetrahedron, i.e.
\be
\mathcal{H}=L^2(G^4/G)=\bigoplus_{j_i\in\frac{\mathbb{N}}{2}}\text{Inv}\left(\otimes_{i=1}^4\mathcal{H}^{j_i} \right),
\ee 
where $\mathcal{H}^{j_i}$ corresponds to the Hilbert space of an irreducible unitary representation of $G=\mathrm{SU}(2)$. The  index $\iota$ labels elements in a basis in $\mathcal{H}$. In this way, it is clear that the presence of the intertwiners with label $\iota_r $ is due to the imposition of the right-invariance onto the field, while the left-invariance leads to the label $\iota_l$, respectively. Hence, the quantum geometric content of the field is stored in the scalar functions $\sigma^{j_1...j_4, \iota_l\iota_r}(\phi)$ in the spin representation

When Eq.~(\ref{meanfielddecomposed}) is injected into the action~(\ref{actioneprl}), one obtains an equation of motion for the condensate field which is a non-linear tensor equation and as such is notoriously difficult to solve, see Refs.~\cite{GFTmin,GFTotherstat,GFTclsol}. We refrain from explicating the full details of this equation in the general case here and direct the reader to the original literature where thez are given in depth~\cite{GFCFriedmann,aniso2}. Instead, we will focus on giving some details of a specific scenario relevant to cosmology the elaboration of which has led to most of the results of the condensate program.

\section{Overview of important results}\label{sec:results}

\subsection{Recovery of Friedmann-like dynamics and bouncing solutions}\label{subsection:resultsA}

What allows to make progress is to focus on the case where the condensate wave function only depends on a single spin variable, as discussed in the following. In fact, this corresponds to an isotropic restriction which leads to a highly symmetric configuration: In this way the condensate is made of equilateral tetrahedra which are the most \enquote{isotropic} configurations in a simplicial context. In effect, the domain of the left- and right-invariant field is reduced to a $1$-dimensional manifold which is parametrized by a single variable, interpreted as the volume and the configuration space is that of a homogeneous and isotropic universe~\cite{GFCFriedmann,aniso2}.\footnote{We comment below on an alternative notion of isotropy which has been explored so far in the literature.  However, notice that such a reduction is a common simplification also applied in the closely related contexts of tensor models for quantum gravity~\cite{TM} and lattice gravity approaches~\cite{QRC,EDTCDT}.} One thus requires the mean field to be of the form
\be
\sigma^{j,\iota\iota}(\phi)=\sigma^{j_1 j_2 j_3 j_4,\iota_l\iota_r}(\phi)\delta^{\iota\iota_l}\delta^{\iota\iota_r}\prod_{i=1}^4\delta^{jj_i}.
\ee
where the identification of intertwiner labels is due to the requirement that the volume be maximized in equilateral tetrahedra~\cite{GFCFriedmann,aniso2}. For such field configurations the action~(\ref{actioneprl}) (when dropping all repeated intertwiner labels for convenience) reads as
\be
S=\int\text{d}\phi\sum_j\bar{\sigma}^{j\iota}\mathcal{K}^{j\iota}\sigma^{j\iota}+\frac{\lambda}{5}\int\text{d}\phi\sum_j\left(\sigma^{j\iota}\right)^5\mathcal{V}_5(j;\iota)~~+~~\text{c.c.}
\ee
with 
\be
\mathcal{V}_5(j;\iota)=\mathcal{V}_5(\underbrace{j,...,j}_{10};\underbrace{\iota,...,\iota}_{5})=f(j;\iota)\omega(j,\iota)
\ee
and
\be
\omega(j,\iota)=\sum_{m_i}\prod_{i=1}^{10}(-1)^{j_i-m_i} \mathcal{I}^{jjjj,\iota}_{m_1 m_2 m_3 m_4}\mathcal{I}^{jjjj,\iota}_{-m_4 m_5 m_6 m_7}\mathcal{I}^{jjjj,\iota}_{-m_7 -m_3 m_8 m_9}\mathcal{I}^{jjjj,\iota}_{-m_9 -m_6 -m_2 m_{10}}\mathcal{I}^{jjjj,\iota}_{-m_{10} -m_8 -m_5 -m_1}.
\ee
The latter product of intertwiners can be cast into the form of a $\{15j\}$-symbol. The details of these calculations can be found in Ref.~\cite{GFCFriedmann} and in greater detail in Ref.~\cite{aniso2}. Again, the specific aspects of the EPRL GFT model would be encoded in the details of the objects $\mathcal{K}^j$ and $\mathcal{V}_5(j;\iota)$ (and thus $f(j;\iota)$). The interaction kernel is supposed to encode the Lorentzian embedding of the theory and thus what is known as the spin foam vertex amplitude with boundary $\text{SU}(2)$-states. Though its details are yet to be put down in the GFT context, its explicit form is not relevant to the results presented below.

With the above, one obtains the equation of motion of the condensate field, i.e.
\be
\mathcal{K}^j\sigma_j(\phi)+\mathcal{V}_{5}^j\bar{\sigma}_j(\phi)^4=0.
\ee
Most generally, the contribution of the kinetic term takes the form $\mathcal{K}^j=A_j\partial_{\phi}^2-B_j$, where $A_j$ and $B_j$ parametrize ambiguities in the EPRL GFT model~\cite{GFCFriedmann,aniso2} and the partial derivatives with respect to the relational clock follow from a derivative expansion with respect to the same variable~\cite{GFCFriedmann,GFTclock}. In what follows, we will see that the requirement that the Friedmann equations be recovered allow to constrain the form of $A_j$ and $B_j$.

\begin{figure}[!h]
\centering%
\begin{minipage}{0.5\textwidth}
  \includegraphics[width=0.5\linewidth]{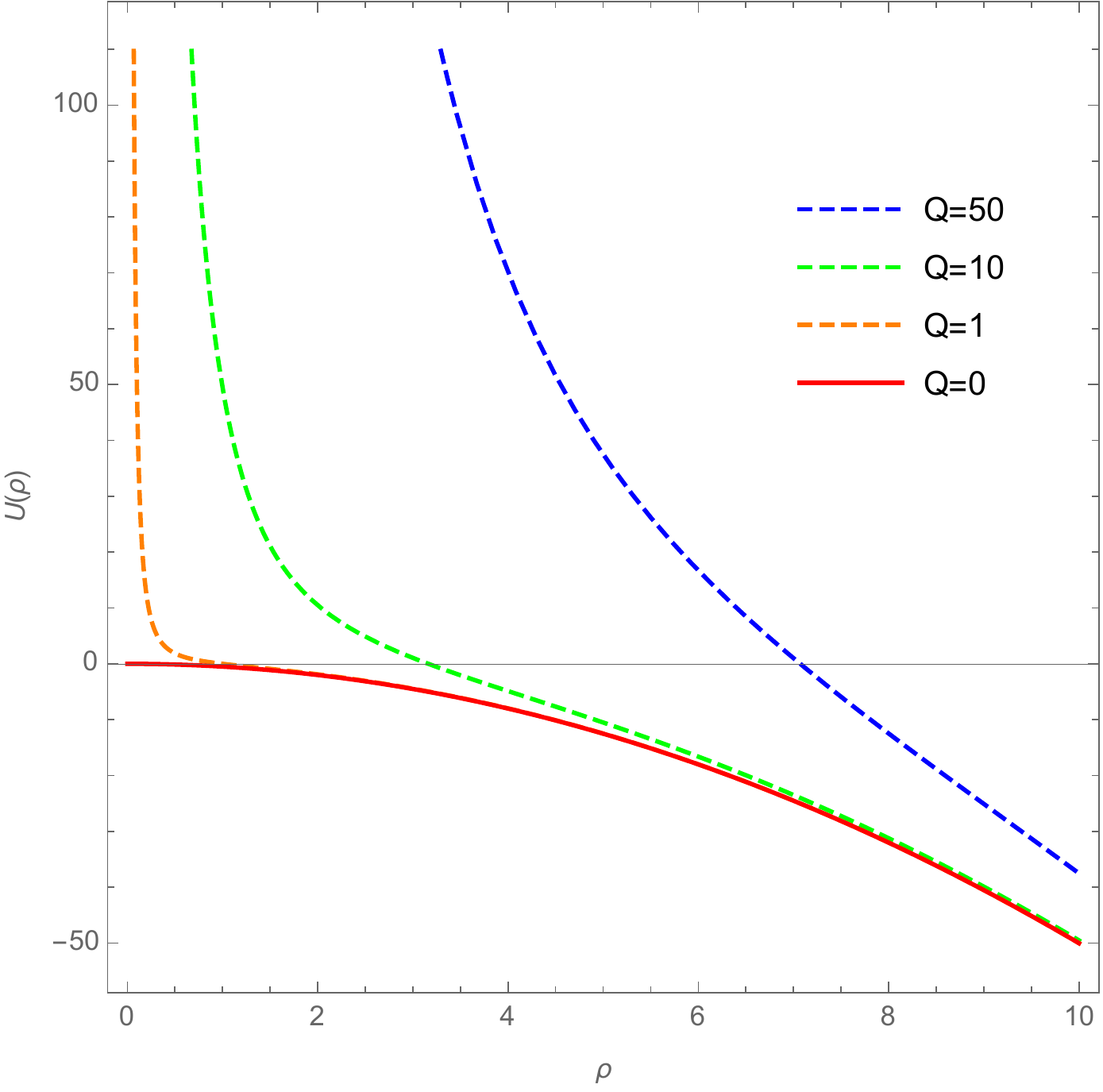}
\end{minipage}%
\begin{minipage}{0.5\textwidth}
  \includegraphics[width=0.5\linewidth]{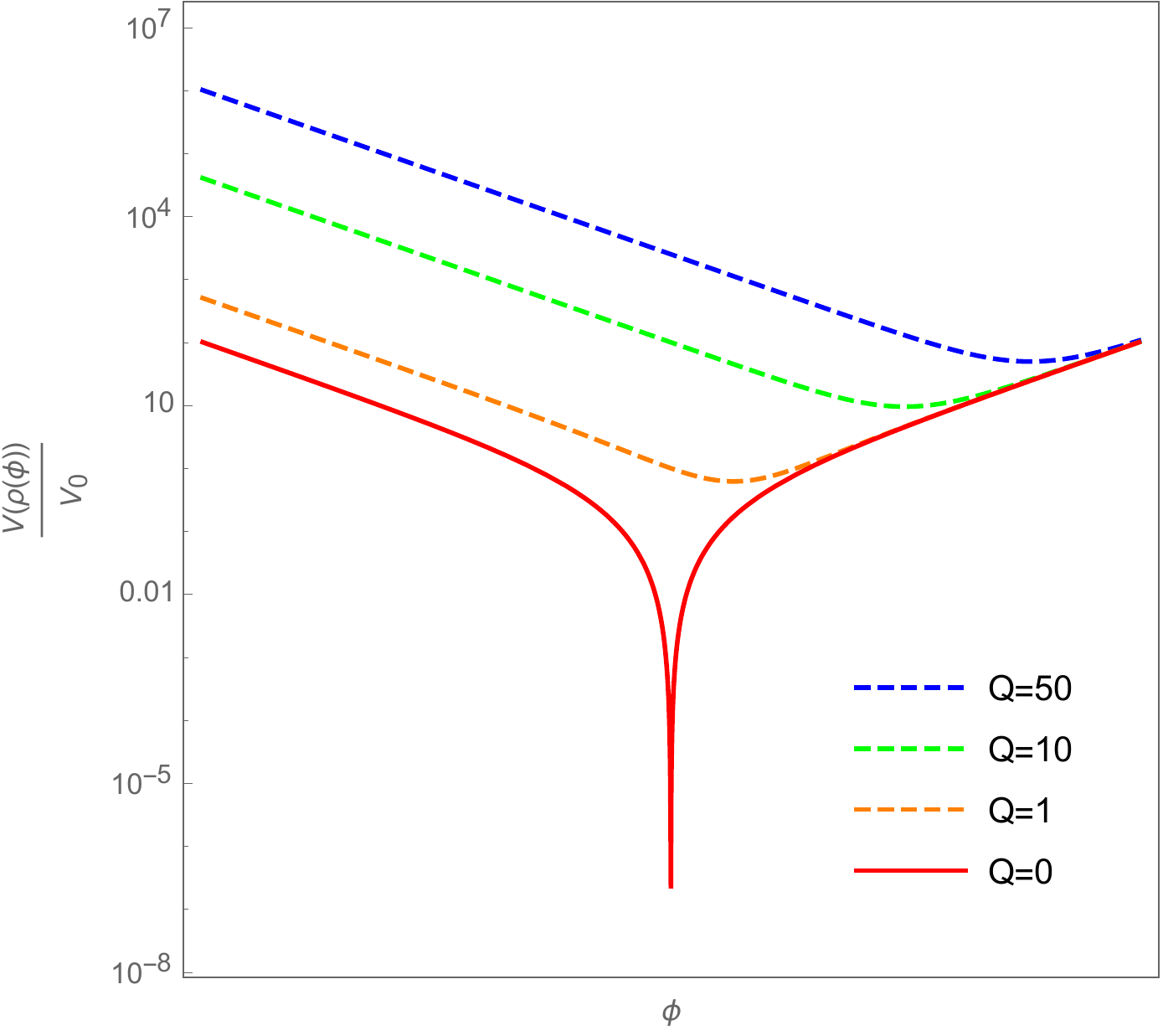}
\end{minipage}%
\caption[bouncefree]{Equation~(\ref{eombounce}) has the form of the equation of motion of a classical point particle with potential $U(\rho)=-\frac{1}{2}m^2\rho^2+\frac{Q^2}{2\rho^2}$. The potential is plotted for different values for $Q$ while $m$ and is kept fixed (left). Solutions to Eq.~(\ref{eombounce}) (initial conditions are arbitrarily chosen) lead to a plot for the volume (right). The point in relational time where the minimum of solutions is reached corresponds to the bounce. The solution for $Q=0$ does not exhibit a bounce since the volume vanishes when $U(\rho)$ turns to zero. In the plots and in this caption the label $j$ is suppressed.}\label{bouncefree}
\end{figure}

To this aim, we follow Refs.~\cite{GFCFriedmann,GFCbounce} and consider the regime of the dynamics where the interaction term is sufficiently small as compared to the kinetic term. Since higher powers of the condensate field are directly proportional to the number of condensate constituents, we may refer to a regime where the interaction term is sub-dominant as being mesoscopic.\footnote{Notice that the term \enquote{mesoscopic} used here only refers to the number of quanta $N$ so far. Detailed studies have to determine the exact range of $N$ for such a regime to hold true and relate it to a range of length scales in the future. Conversely, this would necessitate to study the regimes of very small and very large $N$ where the simple field coherent state ansatz is expected to be inapplicable.} This is a crucial approximation to recover the Friedmann equations below. Then, the equation of motion reduces to
\be\label{eomcondensate}
\partial_\phi^2\sigma_j(\phi)-m_j^2\sigma_j(\phi)=0,~~~\text{with}~~~m_j^2=\frac{B_j}{A_j}
\ee
and when using the polar decomposition of the field as $\sigma_j(\phi)=\rho_j(\phi)\text{e}^{i\theta_j(\phi)}$, yields
\be\label{eombounce}
\rho_{j}''-\frac{Q_j^2}{\rho_j^3}-m_j^2\rho_j=0
\ee
together with the conserved quantities 
\be
Q_j=\rho_j^2\theta'_j~~~\text{and}~~~E_j=(\rho_j')^2+\rho_j^2(\theta_j')^2-m_j^2\rho_j^2.
\ee
Notice that the central term in Eq.~(\ref{eombounce}) diverges towards $\rho_j\to 0$ to the effect that the system exhibits a quantum bounce (elaborated further below), as long as at least one $Q_j$ is non-vanishing, see Fig.~\ref{bouncefree}.
This is the case when requiring the energy density of the clock field to be non-zero. This energy density is given in terms of its conserved momentum $\pi_{\phi}=\sum_j Q_j$ by 
\be
\rho_{\phi}=\frac{\pi_{\phi}^2}{2 V^2}
\ee
where $V$ denotes the expectation value of the volume operator. (Note that $\rho_{\phi}$ is not to be confused with $\rho$ or $\rho_j$ ascribed to the mean field.) It is explicitly given by
\be\label{volumetime}
V(\phi)=\sum_j V_j\rho_j(\phi)^2,~~~\text{with}~~~V_j~\sim j^{3/2}\ell_{\text{Pl}}^3.
\ee
We are now ready to give the dynamics of the volume of the emergent space, namely
\be\label{genfriedmann}
\left(\frac{V'}{3V}\right)^2 = \left(\frac{2\sum_j V_j\rho_j\text{sgn}(\rho'_j)\sqrt{E_j-Q_j^2/\rho_j^2+m_j^2\rho_j^2}}{3\sum_j V_j\rho_j^2}\right)^2~~~\text{and}~~~\frac{V''}{V}=\frac{2\sum_j V_j\left(E_j+2m_j^2\rho_j^2\right)}{\sum_j V_j\rho_j^2},
\ee
as obtained in Ref.~\cite{GFCFriedmann} and call these the generalized Friedmann equations.

The classical limit of these equations is obtained when considering sufficiently large volumes for which the terms with $E_j$ and $Q_j$ in Eqs.~(\ref{genfriedmann}) are suppressed. If one identifies then also $m_j^2\equiv 3\pi G_{\text{N}}$ (where $G_{\text{N}}$ denotes Newton's constant), one recovers the classical Friedmann equations of GR for a flat universe in terms of the relational clock $\phi$,
\be 
\left(\frac{V'}{3V}\right)^2=\frac{4\pi G_{\text{N}}}{3}~~~\text{and}~~~\frac{V''}{V}=12\pi G_{\text{N}}.\footnote{We exemplify the link to the standard Friedmann equations of GR for a flat universe in proper time $t$ as compared to those in relational time $\phi$ via the the first Friedmann equation, i.e.
\be
H^2=\left(\frac{V'}{3V}\right)^2\left(\frac{\text{d}\phi}{\text{d}t}\right)^2~~~\text{with}~~~\pi_{\phi}= V\dot{\phi}.
\ee
The second Friedmann equation can be rewritten in a similar way, see e.g. Ref.~\cite{GFCFriedmann}. This makes transparent that the dynamical equations for the volume as derived by GR and the condensate program take the same form. Notice that the concept of proper time does not exist in the GFT context.}
\ee
Notice that the definition of $G_{\text{N}}$ is understood as a definition in terms of the microscopic parameters $m_j$ (or $A_j$ and $B_j$) and not as an interpretation of the latter.

Another relevant situation where the dynamics of the volume can be solved exactly, is when the condensate is dominated by a single spin $j_o$.\footnote{This is akin to what is done in LQC, where one assumes that the links of the underlying spin network are all identically labeled, with $j=\frac{1}{2}$ being the most studied case~\cite{LQC}. We will present a possible dynamical mechanism leading to a single-spin condensate further below.} In this case the Eqs.~(\ref{genfriedmann}) yield
\be
\left(\frac{V'}{3V}\right)^2 = \frac{4\pi G_{\text{N}}}{3}\left(1-\frac{\rho_{\phi}}{\rho_c}\right)+\frac{4V_{j_o}E_{j_o}}{9V}~~~\text{and}~~~\frac{V''}{V}=12\pi G_{\text{N}}+\frac{2 V_{j_o}E_{j_o}}{V}
\ee
where $\rho_c\sim \frac{3\pi}{2 j_{o}^3}\rho_{\text{Pl}}$ is a critical density~\cite{GFCbounce}. The terms involving $\rho_c$ and $E_{j_o}$ correspond to quantum corrections where the one involving $\rho_c$ is responsible for the quantum bounce. To the past of this event, the emergent space contracts while it expands to the future. It should be remarked that up to the terms depending on $E_{j_o}$ these equations are exactly the modified Friedmann equations derived in LQC.\footnote{In fact, these background dynamics are understood to generalize the effective dynamics of LQC which can be retained as a special case. We refer to Refs.~\cite{GFTevol,GFThamevol} where this point was further explored.}\footnote{In the given picture, the cosmological dynamics expressed by the expansion of the volume is vastly driven by a growing occupation number~\cite{GFCExample2,GFClatticerefinement}. It should be remarked that this is a GFT realization of the lattice refinement of LQC~\cite{LQC}.} For $E_{j_o}>0$ the bounce takes place at an energy density larger than $\rho_c$, while for $E_{j_o}<0$ the bounce is realized for an energy density smaller than $\rho_c$. Independently of the exact value of $E_{j_o}$, a bounce will occur. It should nevertheless be clear that the physical meaning of the conserved quantity $E_j$, from a fundamental point of view, is yet to be clarified. In future research it would also be important to consider the impact of different $j$-modes onto the dynamics. This is in principle straightforward but would then require to solve Eqs.~(\ref{genfriedmann}) numerically. 

Finally, to contextualize, notice that a quantum gravity induced bounce falls into the more general class of bouncing cosmologies which present tentative alternatives to the standard inflationary scenario to resolve the problems of the standard model of cosmology, see Ref.~\cite{Brandenberger} for an overview.\footnote{In Ref.~\cite{GFCmimetic} the relations between the condensate program and mimetic gravity were explored. Mimetic gravity is a Weyl-symmetric extension of GR~\cite{mimeticg} proposed to mimic the effects of cold dark matter within the context of modifications of GR. In the context of limiting curvature mimetic gravity it is possible to realize non-singular bouncing cosmologies in the sense that it is possible to reproduce their background dynamics. This has been shown for the case of LQC~\cite{mukhcham} and very recently for the case of the effective dynamics of GFT condensates~\cite{GFCmimetic}.}\newline

We may list important side-results which support the findings presented above:

\begin{itemize}
\item In Ref.~\cite{GFClowspin}, it is shown that for growing relational time, the condensate dynamically settles into a low-spin configuration, i.e. it will be dominated by the lowest non-trivial representations labeled by $j$. This goes in hand with a classicalization of the emergent geometry~\cite{aniso1}. Following Ref.~\cite{GFClowspin}, this can be seen from the general solutions to Eq.~(\ref{eomcondensate}), i.e.
\be 
\sigma_j(\phi)=\alpha^{+}_j \text{exp}\left(\sqrt{\frac{B_j}{A_j}}\phi\right)+\alpha^{-}_j \text{exp}\left(-\sqrt{\frac{B_j}{A_j}}\phi\right)
\ee
which either lead to exponentially expanding and contracting or oscillating solutions depending on the sign of the argument of the root function. All models for which $B_j/A_j$ has a positive maximum for some $j=j_o$ (as long as $j=0$ is excluded\footnote{We refer to Refs.~\cite{GFTLQG,GFClowspin} for a discussion touching on the subtle differences in between the Hilbert spaces of GFT and LQG especially relevant to the point of the zero-mode $j=0$.}) lead to
\be 
\lim_{\phi\to \pm\infty} V(\phi) = V_{j_{o}} |\alpha^{\pm}_{j_{o}}|^2\text{exp}\left(\pm 2 \sqrt{\frac{B_{j_{o}}}{A_{j_{o}}}}\phi\right).
\ee 
A low-spin configuration is dynamically reached if the maximum of $B_j/A_j$ occurs at a low $j_o$. This was demontrated for reasonable choices of $A_j$ and $B_j$ in Refs.~\cite{GFClowspin} and~\cite{aniso1} to lead to $j_o=\frac{1}{2}$. One may then argue that the type of configuration which is usually assumed in the LQC literature can be derived from the quantum dynamics of GFT.

\item A careful analysis shows that the identification $m_j^2\equiv 3\pi G_{\text{N}}$ only holds asymptotically for large $\phi$, rendering $G_{\text{N}}$ a state-dependent function~\cite{GFCaccel1}. This is illustrated in Fig.~\ref{GN}.  

\begin{figure}[!h]
\centering%
\begin{minipage}{0.5\textwidth}
  \includegraphics[width=0.6\linewidth]{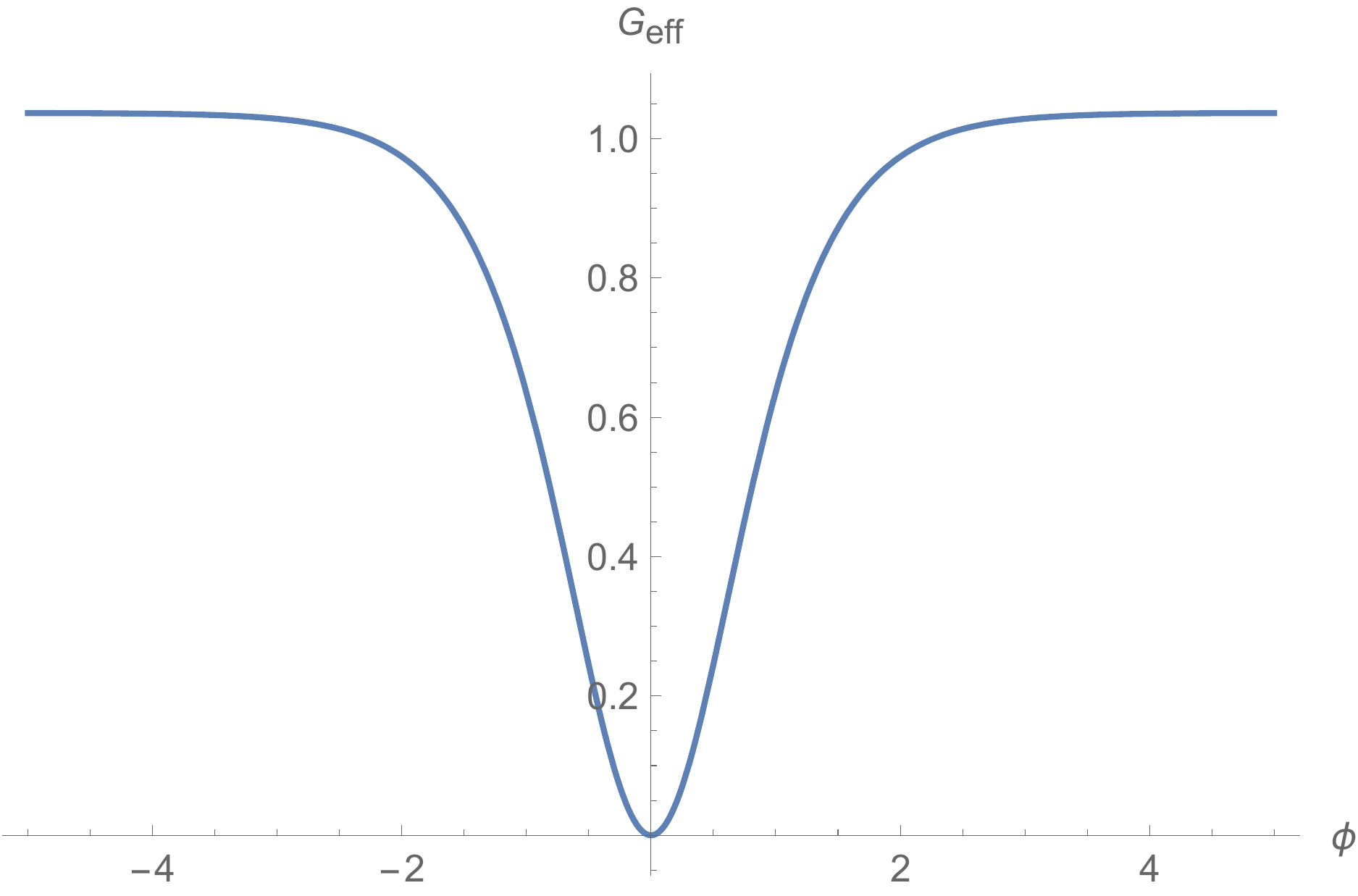}
\end{minipage}%
\begin{minipage}{0.5\textwidth}
  \includegraphics[width=0.6\linewidth]{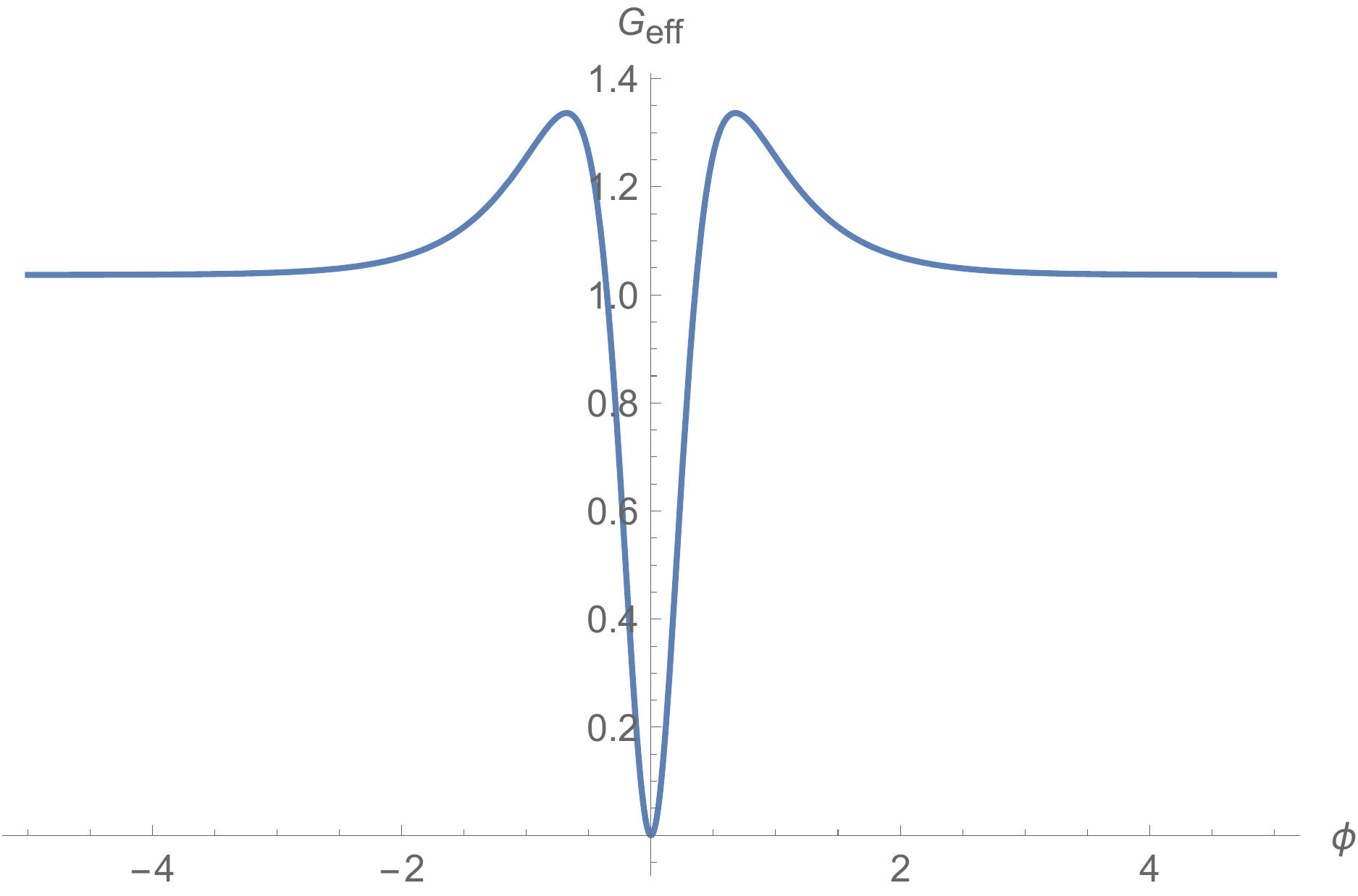}
\end{minipage}%
\caption[effective GNs]{
The effective gravitational constant as a function of
   relational time $\phi$ for $E<0$ to the left and $E>0$
   to the right given in arbitrary units, taken from Ref.~\cite{GFCaccel1}. A bounce occurs towards $\phi=\Phi$ situated at the origin of both plots. For large $\phi$ Newton's constant is asymptotically attained. In the plots and in this caption the label $j$ is suppressed.
}\label{GN}
\end{figure}

\item Another related notion of isotropic restriction has been studied in the literature so far where the condensate is built from tri-rectangular tetrahedra~\cite{aniso1}. This produces physically equivalent results in terms of the dynamics of the volume, as one would expect when invoking naive universality arguments. Notice that both isotropic restrictions correspond to symmetry reductions applied to the quantum state and thus should by no means be equated with those performed in WdW quantum cosmology or LQC. In the latter cases, symmetry reductions are imposed before quantization and this procedure is expected to violate the uncertainty principle~\cite{WdWQC}.\footnote{For a recent attempt at imposing a quantum counterpart of the classical symmetry reduction in the LQG context, we refer to Refs.~\cite{qrLQG}.} In light of the above, it would be important to give a precise notion of isotropy in terms of a properly defined GFT curvature operator.

\item In a related model which does not make use of the relational clock, the field content has been explicitly studied for free and effectively interacting scenarios~\cite{GFCstatic}. For such static configurations one finds that the condensate consists of many GFT quanta residing in the lowest spin configurations. This is indicated by the analysis of the discrete spectra of the geometric operators, as illustrated by Fig.~\ref{volume}. This also supports the idea that under the given isotropic restrictions, such GFT condensate states are suitable candidates to describe effectively continuous homogeneous and isotropic $3$-spaces built from many smallest building blocks of the quantum geometry.

\begin{figure}[!h]
\centering%
\begin{minipage}{0.5\textwidth}
  \includegraphics[width=0.6\linewidth]{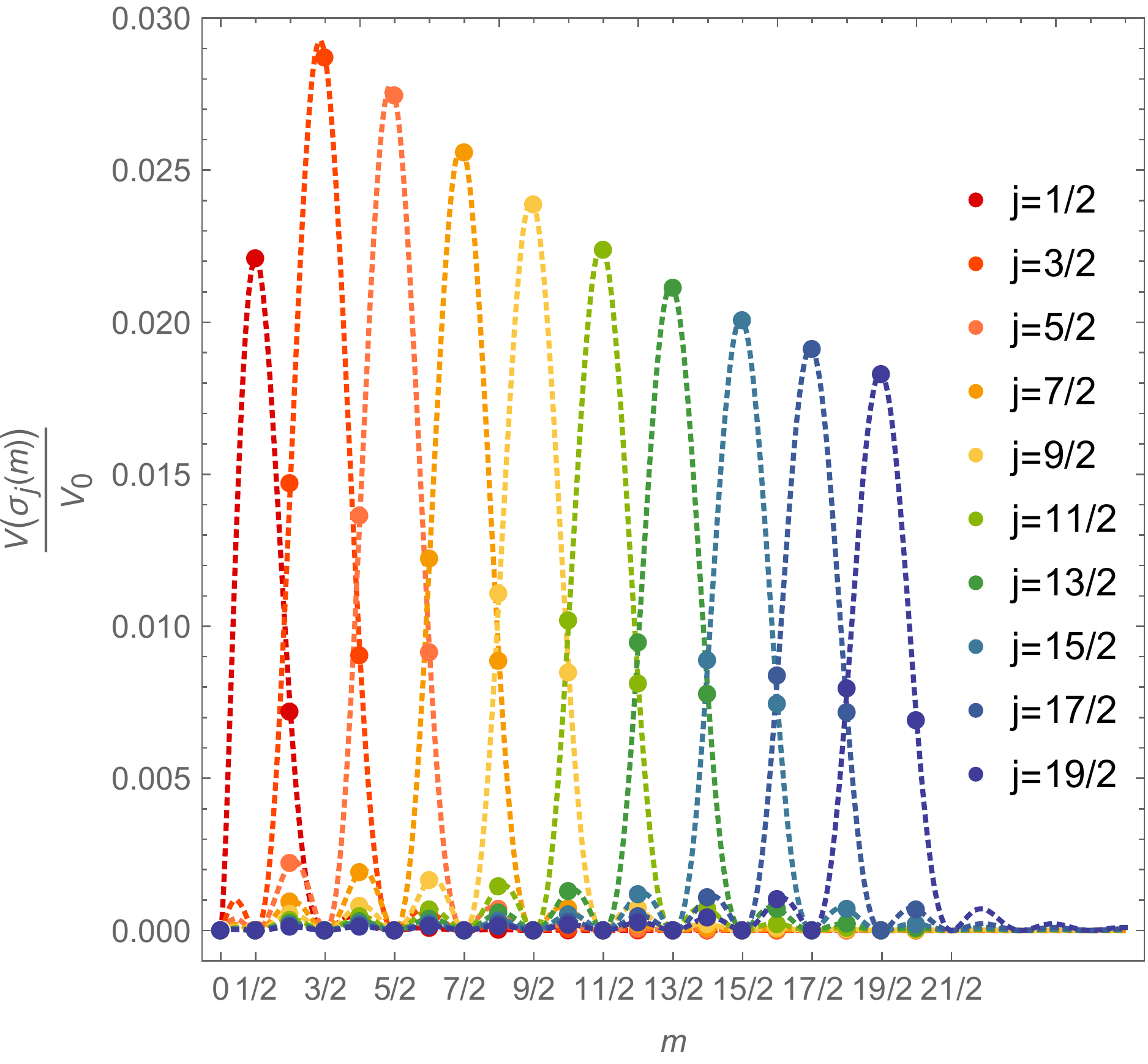}
\end{minipage}%
\begin{minipage}{0.5\textwidth}
  \includegraphics[width=0.6\linewidth]{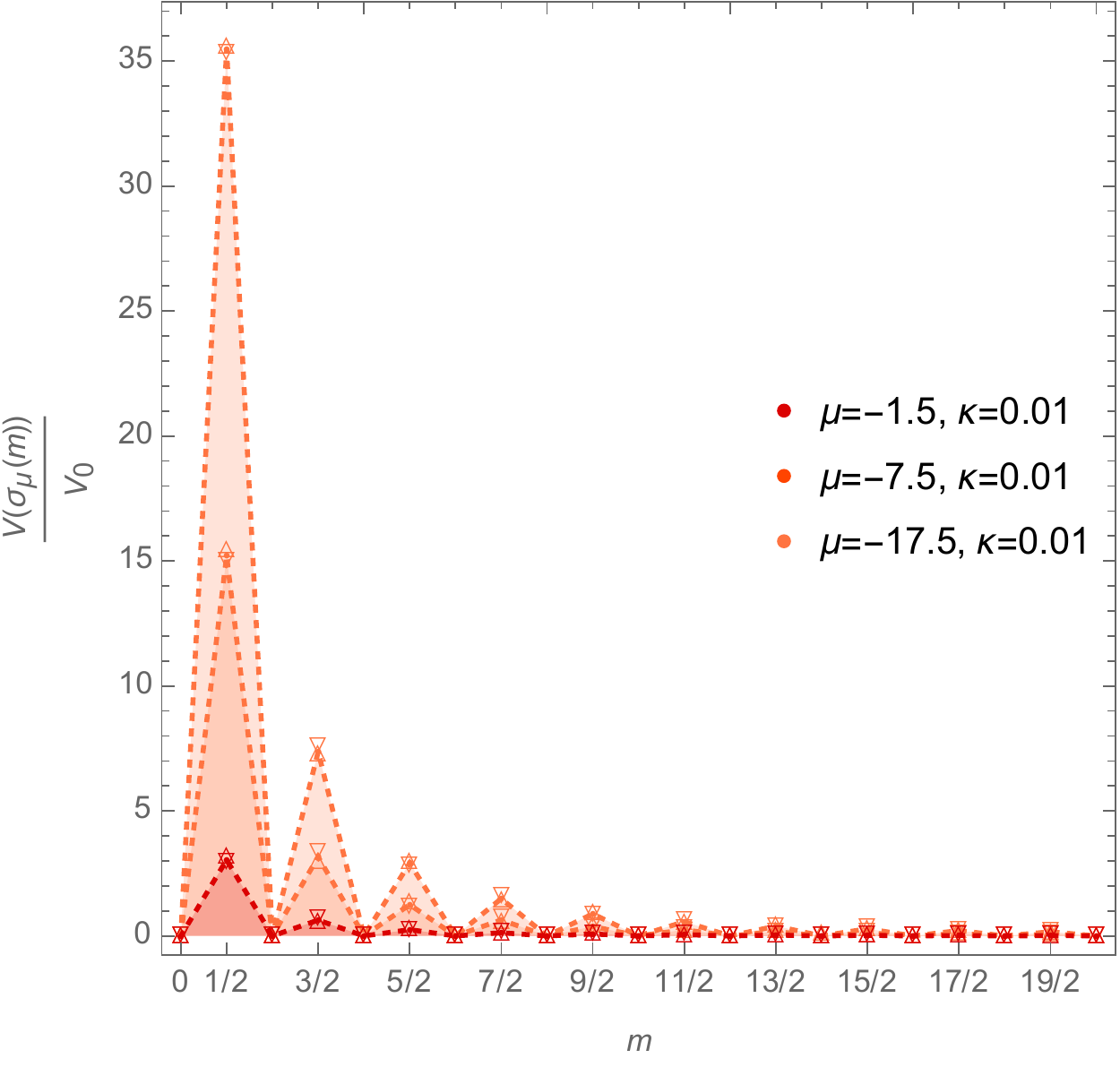}
\end{minipage}%
\caption[spectrum Volume free and interactiong]{The spectrum to the volume operator in the free case to the left and effectively interacting case to the right. These illustrate that the geometric operators are dominated by the lowest non-trivial modes characterizing the condensate field. These plots were taken from Ref.~\cite{GFCstatic} where a detailed discussion of their underlying computations can be found.
}\label{volume}
\end{figure}
\end{itemize}

\subsection{Cyclic cosmologies and accelerated expansion}

In a series of works the effect of simplified GFT interactions onto the cosmological dynamics has been investigated under the assumption that the spin representation $j$ is fixed~\cite{GFCaccel2,aniso1} (which may be motivated by the above-described process of reaching a low-spin configuration). Given these phenomenologically motivated interactions, the equations of motion take a simple non-linear form. Despite the fact that from a GFT point of view such interactions seem to be somewhat artificial due to their lack of a discrete geometric interpretation, they bring us nearer to the physics which we want to probe as they capture the basic non-linearity of the original GFT interactions. The form of the effective potential is given by
\be
\mathcal{V}=B |\sigma(\phi)|^2+\frac{2}{n}w|\sigma(\phi)|^n+\frac{2}{n'}w'|\sigma(\phi)|^{n'},~~~\text{with}~~~n>n'
\ee
and we require $w'>0$ so that the potential is bounded from below. Using the polar form of the field, we obtain the equation of motion
\be\label{eomeffint}
\rho''-\frac{Q^2}{\rho^3}-m^2\rho+\lambda\rho^{n-1}+\mu\rho^{n'-1}=0
\ee
where we have set 
\be
\lambda \equiv -\frac{w}{A}~~~\text{and}~~~\mu\equiv-\frac{w'}{A}.
\ee
To guarantee that this equation does not lead to an open cosmology expanding at a faster than exponential rate, we have $\mu>0$. In consistency with the free case discussed above, $m^2>0$ while the sign of $\lambda$ can be left unconstrained. A first observation from this equation of motion is its resemblance with that of a classical point particle in the potential
\be 
U(\rho)=-\frac{m^2}{2}\rho^2+\frac{Q^2}{2\rho^2}+\frac{\lambda}{n}\rho^n+\frac{\mu}{n'}\rho^{n'}
\ee
so that with the given signs and the bouncing contribution of strength $Q^2$ the solutions to Eq.~(\ref{eomeffint}) yield cyclic motions. Via Eq.~(\ref{volumetime}) these correspond to cyclic solutions for the dynamics of the emergent universe. Hence, we observe that bounded interactions induce a recollpase. Given that in the classical theory a recollapsing solution follows from a closed topology of $3$-space, this might give an indication of how to obtain such topologies from these simple GFT condensates. 

\begin{figure}[!h]
\centering%
\begin{minipage}{0.5\textwidth}
  \includegraphics[width=0.6\linewidth]{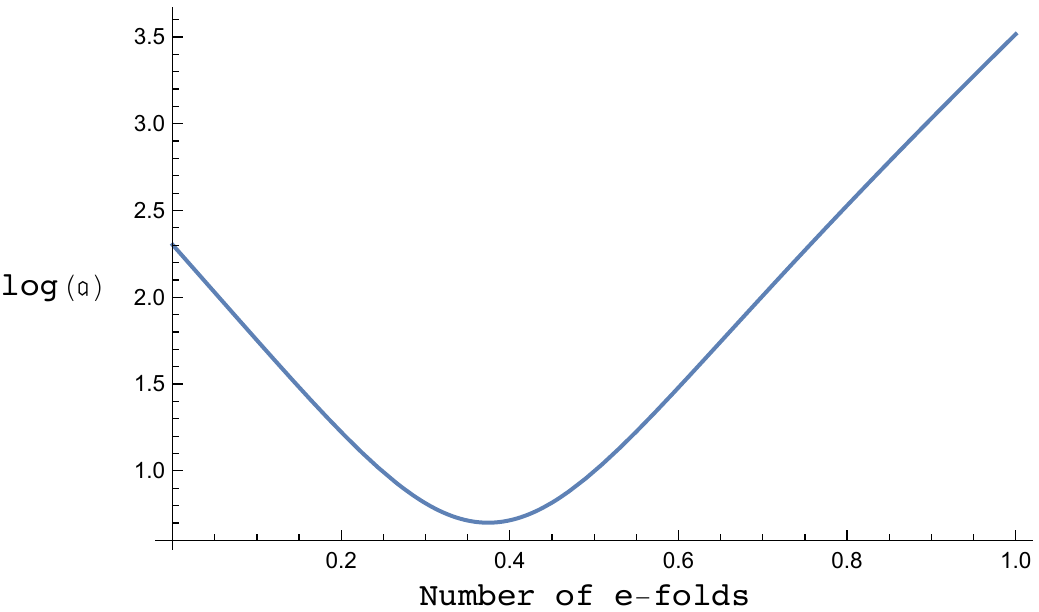}
\end{minipage}%
\begin{minipage}{0.5\textwidth}
  \includegraphics[width=0.6\linewidth]{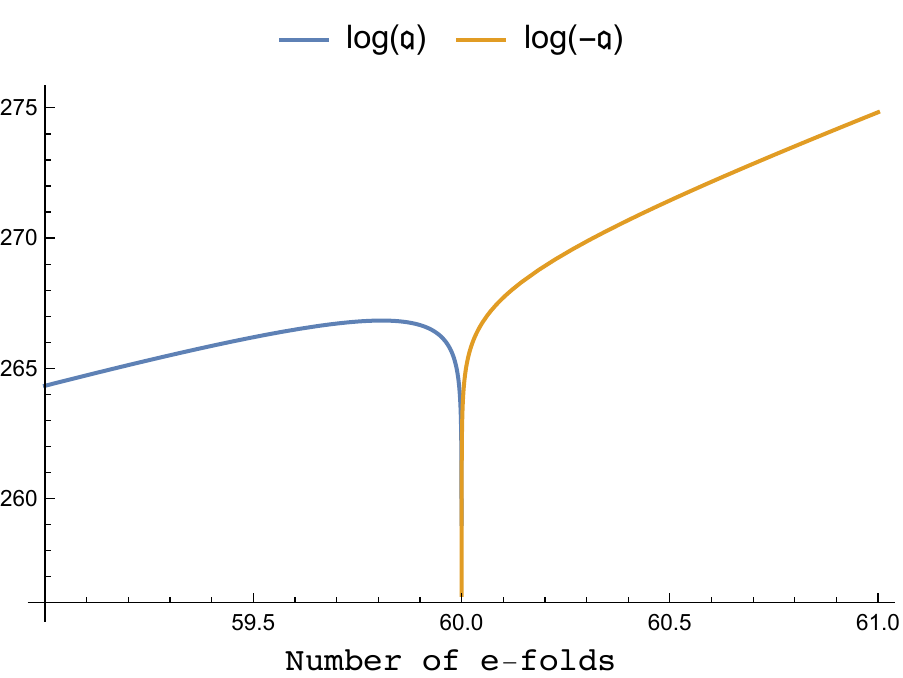}
\end{minipage}%
\caption[efolds eff interactions]{The plot to the left shows the behavior of the acceleration close to the bounce while the one to the right illustrates it towards the end of inflation. These plots were taken from Ref.~\cite{GFCaccel2} to which we direct for details.}\label{efolds}
\end{figure}

Regarding the expansion behavior of the emergent geometry, using the above-given interactions it is possible to obtain a long lasting accelerated phase after the bounce. In fact, the free parameters may be fine-tuned to achieve any desirable value of $\text{e}$-folds so that this behavior can be understood as an inflationary expansion of quantum geometric origin. This becomes transparent when writing for the number of $\text{e}$-folds
\be
N=\frac{1}{3}\text{log}\left(\frac{V_{\text{end}}}{V_{\text{bounce}}}\right)=\frac{2}{3}\text{log}\left(\frac{\rho_{\text{end}}}{\rho_{\text{bounce}}}\right)
\ee
and incorporating it in an expression for the acceleration. Since there is no notion of proper time in GFT, a sensible definition of acceleration can only be given in relational terms. In particular, we seek a definition that agrees with the standard one given in ordinary cosmology via the Raychaudhuri equation which allows us to define the acceleration as
\be
\mathfrak{a}(\rho)=\frac{V''}{V}-\frac{5}{3}\left(\frac{V'}{V}\right)^2
\ee
the derivation of which is discussed in detail in Refs.~\cite{GFCaccel1,GFCaccel2}. Using this, one finds that the free case does not lead to a value of $N$ large enough to supplant the standard inflationary mechanism in cosmology. However, the careful analysis of Ref.~\cite{GFCaccel2} demonstrates that with a hierarchy $\mu\ll |\lambda|$ together with $\lambda>0$ leads to $n'>n\geq 5$ which allows room for an era of accelerated expansion analogous to that of models of inflationary cosmology. If phantom energy is ruled out, only $n=5$ and $n'=6$ are admissable selecting an interaction term which is in principle compatible with simplicial interactions, introduced above. These results are illustrated in Fig.~\ref{efolds}. Notice that these works emphasize the role that phenomenology can take for model building in quantum gravity. 

It should be emphasized that these findings have a purely quantum geometric origin and in particular are not based in any way on the assumption of a specific potential for the minimally coupled massless scalar field $\phi$, the relational clock. This is in stark contrast to inflation which depends on the choice of potential and initial conditions of the inflaton field to yield the desired expansion behavior~\cite{Mukhanov,Linde,schism}.

\subsection{Anisotropies and inhomogeneities}

If quantum gravity is to offer the picture of the earliest moments of our Universe, it must include an approximately homogeneous and isotropic background with superimposed perturbations. Given the above results an important step for the condensate cosmology program is to go beyond the considered isotropic restriction and homogeneous configurations and to study more general configurations and their dynamics. In the following, we want to briefly discuss recent advances in which the exploration of anisotropies~\cite{aniso1,aniso2} and inhomogeneities~\cite{GFCinhom} has been commenced.\newline 

The study of anisotropic GFT configurations and their dynamics is of general importance since it would be desirable to see if at least a subset of these can be in agreement with the observed isotropy of our Universe at late times. For these one has to show that anisotropies do not grow in the expanding phase. Apart from that, it is well known that bouncing cosmologies are haunted by the notorious problem of uncontrolled growth of anisotropies when the universe contracts~\cite{Brandenberger}. This is the problem of the Belinsky-Khalatnikov-Lifshitz instability~\cite{BKL}. In light of this, it is interesting to understand the fate of anisotropies when approaching the quantum bounce as predicted by GFT.

\begin{figure}[!h]
\centering%
\begin{minipage}{0.5\textwidth}
  \includegraphics[width=0.6\linewidth]{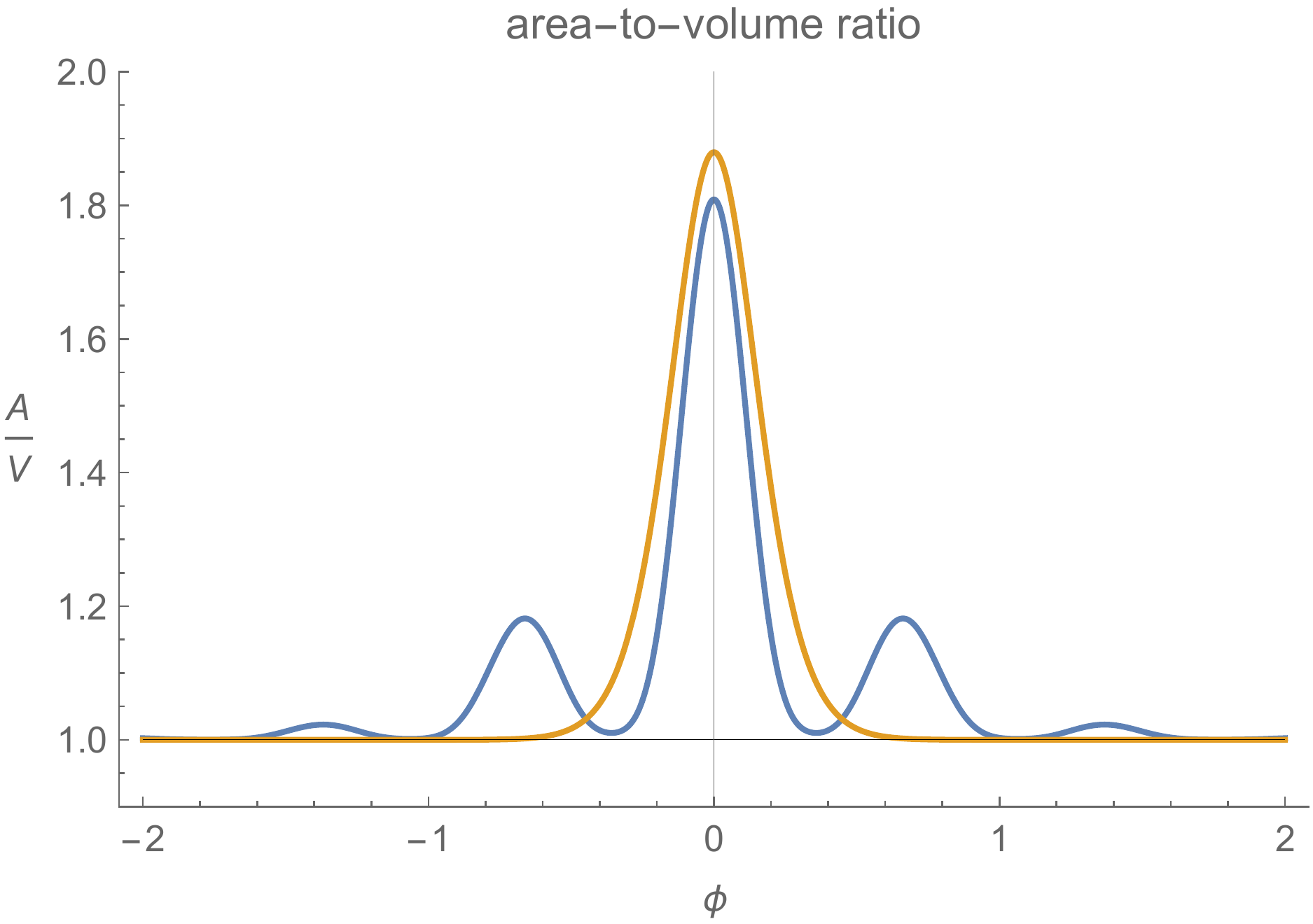}
\end{minipage}%
\begin{minipage}{0.5\textwidth}
  \includegraphics[width=0.6\linewidth]{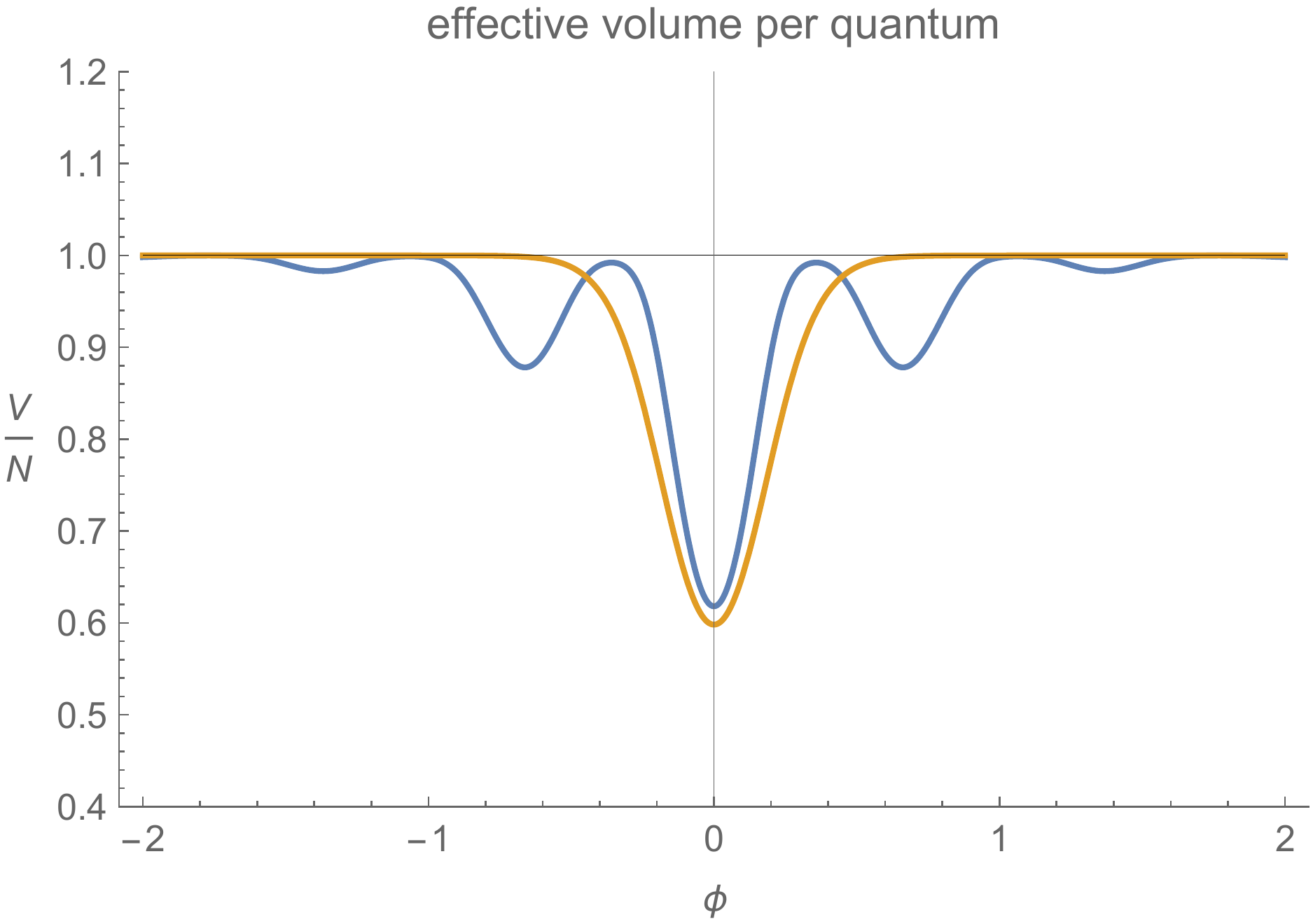}
\end{minipage}%
\caption[degenerate geometry]{The surface-area-to-volume ratio (left) and the effective volume per quantum (right) as a functions of relational time. These figures demonstrate the degenerate character of the quantum geometry towards the bounce and the damping of the anisotropies in the outgoing phase. The plots were taken from Ref.~\cite{aniso2} where detailed explanations are given.}\label{degenerategeometry}
\end{figure}

Leaving the technical details aside, in Ref.~\cite{aniso1} it has indeed been shown for rather general configurations that they dynamically isotropize in relational time by means of a simple mechanism (which is akin to the one responsible for settling the system into a low-spin configutation, as described in Section~\ref{subsection:resultsA}). Conversely, it is demonstrated that anisotropic contributions to the condensate become more and more pronounced towards small volumes. This paved the way to a systematic investigation of anisotropic perturbations over an isotropic background in the vicinity of the bounce in Ref.~\cite{aniso2}. In particular, a region in the parameter space is identified such that these anisotropies can be large at the bounce but are fully under control. From this it also follows that towards the bounce the quantum geometry of the emergent universe is rather degenerate. Furthermore, this analysis shows that the anisotropic perturbations become negligible the further away the system is from the bouncing phase and can be completely irrelevant to the dynamics before interactions kick in. Hence, after the bounce a cosmological background emerges the dynamics of which can again be cast into the form of the above-given effective Friedmann equations, thus corroborating the results of Refs.~\cite{GFCFriedmann,GFCbounce,aniso1}. These results are illustrated in Fig.~\ref{degenerategeometry}. On more general grounds, these studies form a crucial starting point towards identifying anisotropic cosmologies, i.e. Bianchi models, within this approach and allow to establish contact to corresponding studies in WdW~\cite{WdWQC}, spin foam~\cite{SFC,SFC1,SFC2} and loop quantum cosmology~\cite{LQC}.

Modern Cosmology teaches us that the seeds for structure formation are represented by inhomogeneities in the very early Universe~\cite{structure}. Hence, the identification and study of cosmological inhomogeneities in the condensate approach is mandatory to promote it to a realistic contestant theory of quantum cosmology. In particular, the goal would be to find a mechanism rooted in quantum geometry which explains the origins of inhomogeneities without referring to the inflationary paradigm where the inhomogeneities correspond to quantum fluctuations of the inflaton field~\cite{Mukhanov}.\newline

Recent progress building on Refs.~\cite{GFCothers}, allows to extend the formalism beyond homogeneity. In Refs.~\cite{GFCinhom,GFTrods} the formalism of including a free massless scalar field~\cite{GFCFriedmann,GFTclock} is extended to incorporate four reference scalar fields which are used as relational clocks and rods, i.e., as a physical coordinate system. Again, this procedure is common in classical and quantum gravity approaches~\cite{rods,dustlqg}. In this setting quantum fluctuations (i.e. small inhomogeneities) of the local $3$-volume around a nearly homogeneous background geometry are studied. Their power spectrum can be calculated and this is shown to be approximately scale invariant (where the scale is defined by the reference matter), the amplitude is small and decreases as the emergent universe expands. However, it was also demonstrated that analogous statements do not hold for perturbations in the total density of the scalar fields when the gradient energy is non-negligible. Notice that the details of these arguments relied on the specific choice of condensate state which solves the condensate dynamics and thus depends on the approximation scheme summarized in Sections~\ref{condensatestates} and~\ref{condensatedynamics}. More recent work~\cite{GFCinhom2} has shown how the transition from the initial quantum fluctuations present in the deep quantum gravity regime to classical observable inhomogeneities can be accomplished. By and large, it is striking that features of the spectrum of cosmologically relevant observables can be recognized using the condensate formalism. Future rearch has to bridge the gap between observations of the early Universe and the condensate formalism and the hope is that the incorporation of more complicated matter dynamics can reproduce observationally viable results.\footnote{Notice that this proposal of incorporating inhomogeneities has recently been further developed using the separate universe approach to describe long-wavelength scalar perturbations~\cite{sepuniv}.}

\section{Discussion and outlook}\label{sec:discussion}

In this brief review we wanted to draw attention to key results of the GFT condensate cosmology program which illustrate its potential to provide a quantum gravitational foundation for early universe cosmology. Finally, we would like to point to open directions (if not already stated in the main body of this appetizer) and address some relations with other non-perturbative discrete quantum gravity approaches trying to extract cosmological solutions from their path integral formulations.

Beyond its application to cosmology showing a rich phenomenology, this program has revealed an interesting and powerful perspective on the extraction of continuum information from a discrete geometric setting: The field theoretic setting of GFT and specifically the use of field coherent states prove extremely useful and elegant to this aim.

To contextualize this, we may compare with the other non-perturbative and discrete path integral approaches to quantum gravity. In EDT, CDT, tensor models, the discreteness of geometry is regarded as a mathematical tool allowing us to rewrite the continuum path integral in a discrete form. The philosophy behind taking the continuum limit is rather similar among them and the goal is to study the phase structure of the respective theories via analyticity properties of the partition function. The CDT approach has been able to produce physically relevant, i.e. extended macroscopic geometries which obey effective minisuperspace dynamics~\cite{EDTCDT}. This potentially highlights the role of causality in facilitating the escape from the sector of unphysical continuum geometries EDT~\cite{EDTCDT} and TMs~\cite{TM} are so far stuck with. 

A point of criticism often invoked regarding these approaches, concerns the lack of a clear interpretation of expectation values of observables rigorously defined on a physical Hilbert space, which is in principle available in covariant LQG and GFT. Given this, GFT condensate cosmology is not the only approach which tries to extract the cosmological sector of LQG from a covariant formulation of its dynamics.  In the spin foam cosmology approach one uses the spin foam expansion which is an expansion in terms of the number of degrees of freedom~\cite{SFC}. It has mostly been studied for so-called dipole graphs (corresponding to the simplest cellular decomposition of the $3$-sphere)~\cite{SFC1} and can be extended to more general regular graphs~\cite{SFC2}. A central assumption of this approach is that a \textit{fixed} number of quanta of geometry captures all relevant physics. In contrast, in the condensate program one looks for continuum physics away from the Fock vacuum and does not restrict the number of quanta which can be \textit{rather large and dynamical}. Although GFT allows to study an infinite class of simplicial complexes by construction, it provides the field theoretic approximation tools to study the physics of many LQG degrees of freedom while bypassing the treatment of highly complicated spin networks. In addition, in the spin foam context one typically studies the semi-classical limit by requiring the configurations to peak on some triangulated classical geometry~\cite{LQG,covlqg}.\footnote{A different and interesting take on regaining the continuum in spin foam models is presented by the spin foam coarse graining and renormalisation proram for which we refer to Refs.~\cite{SFcg}.} This point of view is also not assumed in GFT condensate cosmology, motivated by the condensate hypothesis which fixes the states to be described by the simple condensate wave function.
  
Being aware that the choice of simple trial states and the disregardance of proper simplicial interactions so far neglects all the connectivity information of those spin networks that would be considered important for a realistic definition of a non-perturbative continuum vacuum state, it is pressing to go beyond the given simplifications. The exploration of the phenomenologically motivated interactions presented here as well as the study of dipole condensates~\cite{GFC} goes into this direction. Notice that recent work~\cite{GFTmin} in the context of the dynamical Boulatov model which is a model for Euclidean quantum gravity in $3d$ has shown that non-trivial condensate solutions can be produced where simplicial interactions are fully considered. The background quantum geometries one yields in this way have to be better understood but this procedure could in principle be carried over to the case in $4d$. Also, it is clear from this example that the choice of simple states does not pose a major problem since more complicated ones associated to connected graphs are then easily generated by the simplicial interaction term. In light of this, it would be important to study the relational evolution of such properly interacting condensates in order to see if the intriguing results regarding the accelerated expansion of the emergent geometry found via exploring the simplified interactions can be reproduced. Furthermore, studying the effect of the simplicial interaction of the Lorentzian EPRL GFT model (or any related model) onto the condensate will require to put it explicitly down in terms of its boundary data~\cite{aniso2}. It could also be interesting to consider a colored version of such a model, given the insight that in perturbative expansion of the partition function the color degree of freedom guarantees that all terms are free of topological pathologies~\cite{coloring}.
 
Apart from their impact on the dynamics, a better understanding of interactions will also allow to construct more sophisticated observables capturing curvature and cosmological anisotropies which in turn will prove indispensable to classify different emergent geometries from one another. It is nevertheless remarkable, that in the regime where interactions are sub-dominant and which has been explored most so far, rich Friedmann-like dynamics can be obtained. 

A related point to be focused on, touches on higher order corrections to the so-far considered condensate equation of motion and to understand if they can be neglected or if they would have a drastic impact on the cosmological interpretation of this approach. Understanding the full quantum dynamics will then shed light onto the phase structure of interesting GFT models of $4d$ quantum gravity. In other words, it has to be checked by means of non-perturbative techniques if a simplicial GFT model for $4d$ quantum gravity can truely exhibit a phase or phases which are related to $(3+1)$-dimensional Lorentzian continuum geometries. In the context of exploring the notion of phases in GFT, it would also be important to understand the relation in between a potential phase transition into a geometric phase for which the order parameter should vanish and the occurrence of a bounce which (by definition) forbids a zero-volume state. 

At the very end, this approach will be judged by the ability to extract phenomenological signatures to see if it can be a realistic contestant theory of quantum cosmology. The development of a scenario to explain the origin of cosmological perturbations neither by the mechanism provided by inflation nor as in ordinary bounce models but rather via the quantum fluctuations of the geometry itself as given by GFT condensates is an important step into this direction.

{\bf Acknowledgements.} The authors would like to thank the two anonymous referees for their remarks which led to an improvement of the manuscript.

\end{document}